\documentclass{article}
\usepackage{graphicx}
\usepackage{amsmath}
\usepackage{amssymb}
\usepackage{authblk}
\usepackage[utf8]{inputenc}
\usepackage[english]{babel}
\usepackage[shortlabels]{enumitem}
\usepackage{algorithm2e}
\usepackage{dsfont}

\newtheorem{theorem}{Theorem}[section]

\newtheorem{lemma}[theorem]{Lemma}
\begin{document}

\title{Approximating Boolean Functions with Disjunctive Normal Form}
\author{Yunhao Yang\thanks{Email: yunhaoyang234@utexas.edu}}
\author{Andrew Tan\thanks{Email: andrewtan@utexas.edu}}
\date{24 April 2020}
\affil[1]{\small{Department of Computer Science, University of Texas at Austin}}
\maketitle

\begin{abstract}
The theorem states that:
Every Boolean function can be \\$\epsilon$ -approximated by a Disjunctive Normal Form (DNF) of size $O_{\epsilon}(2\textsuperscript{n}/log\emph{n})$.
This paper will demonstrate this theorem in detail by showing how this theorem is generated and proving its correctness. 
We will also dive into some specific Boolean functions and explore how these Boolean functions can be approximated by a DNF whose size is within the universal bound $O_{\epsilon}(2\textsuperscript{n}/log\emph{n})$. The Boolean functions we interested in are:

\begin{itemize}
    \item Parity Function: the parity function can be $\epsilon$-approximated by a DNF of width $(1 - 2\epsilon)n$ and size $2^{(1 - 2\epsilon)n}$.
    Furthermore, we will explore the lower bounds on the DNF's size and width.
    
    \item Majority Function: for every constant $1/2 < \epsilon < 1$, there is a DNF of size 2\textsuperscript{$O(\sqrt{n})$} that can $\epsilon$-approximated the Majority Function on n bits.
    
    \item Monotone Functions: every monotone function \emph{f} can be \\$\epsilon$-approximated by a DNF \emph{g} of size $2^{n - \Omega\epsilon(n)}$ satisfying \emph{g}(x) <= \emph{f}(x) for all x.
    
\end{itemize}

\end{abstract}

\section{Universal Bounds}
\newtheorem{definition}{Definition}[section]
\begin{definition}{Disjunctive Normal Form:}
a canonical normal form of a logical formula consisting of a disjunction (OR) of conjunctions (AND).
\end{definition}

The Lupanov's Theorem states that every Boolean function with n variables can be computed by a DNF with size $2^{n-1}$ and width $n$. Then, there is a question about whether we can find a DNF with smaller size that can compute most of the input correctly. In another word, we want to use a DNF to approximate other Boolean functions.

\begin{definition}{$\epsilon-close$:}
The functions $f, g: {0, 1}^{n} \rightarrow {0, 1}$ are $\epsilon-close$ if $|{x\in {0, 1}^{n}: f(x) \neq g(x)}| \le \epsilon2^{n}$
\end{definition}

\begin{definition}{$\epsilon-approximate$:}
A DNF $\epsilon-approximates$ to $f: {0, 1}^{n} \rightarrow {0, 1}$ if the function it computes is $\epsilon-close$ to f.
\end{definition}

We are interested in the universal upper bounds of Disjunctive Normal Form for approximating any Boolean function. It is a strong argument since there are many different kinds of Boolean functions, and this theorem is applicable to all of them. There definitely are many special Boolean functions that have tighter upper bounds, which we will discuss in Section 2, 3 and 4. The following theorem gives a tight universal upper bound for all Boolean functions.

In order to compute the optimal upper bound, they constructed a two-stage process and ensure that both states happen with high probability by choosing the appropriate parameters, as shown in Theorem 1. In the first stage, the algorithm selects a random subset S of $f^{-1}(0)$, and define a function g which equals 1 on every input in $f^{-1}(1)\bigcup S$. The second stage selects a random subset of sub-cubes that are 1-monochromatic in g. The union of sub-cubes corresponds to a DNF that computes a function h. When S is small enough, function h is close to f, and elements in $f^{-1}(1)$ are covered by those sub-cubes.

\begin{theorem}
Let $\epsilon >= 10/n$. Every Boolean function $f : \{0, 1\}^{n} \rightarrow \{0, 1\}$ can be $\epsilon$-approximated by a DNF of size $4\ln(4/\epsilon) * 2^{n-d}$ where:
\begin{center}
    $d = \log\log_{2/\epsilon}(n/(\ln(4/\epsilon)\log\log_{2/\epsilon}(n))$
\end{center}
Which means, every f can be $\epsilon$-approximated by a DNF of size $O_{\epsilon}(2\textsuperscript{n}/log\emph{n})$
\end{theorem}

Now, let's start the proof of Theorem 1: 

Assume $\min {Pr[f(x) = 0], Pr[f(x) = 1]} \ge \epsilon$. Let $g: \{0,1\}^{n} \rightarrow \{0,1\}$ be the random functions that g(x) = 1 if $x \in f^{-1}(1)$. Then, for each $x \in f^{-1}(0)$, set g(x) = 1 with probability $\epsilon / 2$. Let G denote the induced distribution over all Boolean functions and apply the Chernoff bound we can get:
\begin{equation}
    Pr_{G}[Pr_{f^{-1}(0)}[f(x) \neq g(x)] \ge \epsilon] \le e^{-\epsilon^{2}*2^{n} / 3}
\end{equation}

\begin{definition}{Special:}
Let C be a sub-cube, C is special if C has dimension exactly d and the d free coordinates of c are ${dk + 1, ..., dk + d}$ for some $k = 0,...,\lfloor n/d\rfloor - 1$.
\end{definition}

By definition, Special sub-cubes would satisfy the following properties:
\begin{itemize}
    \item there are $\lfloor n/d\rfloor*2^{n - d}$ special sub-cubes
    \item each sub-cube is included with probability $(\epsilon/2)^{2^{d}}$
\end{itemize}

Let $h: \{0,1\}^{n} \rightarrow \{0,1\}$ be the union of a random subset of the \\1-monochromatic special sub-cubes in g and each special sub-cube in g is included in h with probability $(\epsilon / 2)^{\lambda}$, $\lambda$ = number of x in sub-cube C such that f(x) = 1. Note that:
\begin{itemize}
    \item h is a DNF of width n - d.
    \item $h^{-1}(1) \in g^{-1}(1)$.
    \item the error of h on the 0-inputs of f is less or equal to the error of g and Equation \theequation will remain true if we replace g with h:
\end{itemize}

Let $x \in f^{-1}(1)$, the probability of h(x) = 0 equals to the probability that none of the special sub-cubes containing x are included in h. Since any two special sub-cubes only intersect at x, so:

\begin{equation}
    Pr_{G}[h(x) = 0] = (1 - (\epsilon / 2)^{2^{d}})^{\lfloor n/d\rfloor} \le \exp{-(\epsilon/2)^{2^{d}}n/d} < \epsilon/4
\end{equation}
This shows:
\begin{equation}
    E_{G}[Pr_{f^{-1}(1)}[f(x) \neq h(x)]] < \epsilon/4
\end{equation}
and therefore
\begin{equation}
    Pr_{G}[Pr_{f^{-1}(1)}[f(x) \neq h(x)] \ge \epsilon] \le 1/4
\end{equation}

Then, from the properties of special sub-cubes we can get the following:
\begin{equation}
    E_{G}[DNF-size[h]] = (\epsilon/2)^{2^{d}} \lfloor n/d\rfloor*2^{n - d} \le 2\ln(4/\epsilon) * 2^{n - d}
\end{equation}
\begin{equation}
    Pr_{G}[DNF-size[h]\ge 4\ln(4/\epsilon) * 2^{n - d}] \le 1/2
\end{equation}
From Equation 1, 4 and 6, combine the result by union the bounds, we can conclude that
\begin{quote}
    there exists a function h such that $DNF-size[h] \le 4\ln(4/\epsilon) * 2^{n - d}$, and $Pr[f(x)\neq h(x)] \le \epsilon$
\end{quote}
Complete proving Theorem 1.
\newline
Furthermore, there is a more intuitive version of Theorem 1.
\begin{theorem}
Every function f can be 0.1-approximated by a DNF of size less or equal to $ 2^{n}/\log (n)$
\end{theorem}

To prove Theorem 2, first we can flip each 0-input to 1 independently with probability $\epsilon/2$. On this condition,error on 0-inputs is less or equal to $\epsilon$. Second, let $d = \log\log(n)$, partition $[n]$ into $n/d$ blocks of size d. Every x is contained in $n/d$ special sub-cubes. So, we have

\begin{equation}
    Pr[\text{x is not covered}] = (1 - \epsilon^{2^{d}})^{n/d} \le \epsilon / 4
\end{equation}
Third, note that each special sub-cube included with probability exactly $\epsilon^{2^{d}}$. Therefore,
\begin{equation}
    Pr[\lambda] = \epsilon^{2^{d}}*n/d*2^{n-d} \approx 2^{n}/\log (n)
\end{equation}
Proved.

\section{Approximating Parity Function with DNF}
We are interested in whether the universal bound we showed above can apply to every Boolean functions. First, we choose Parity function and compute the upper bound of size of the DNF which approximates the Parity function.
\begin{definition}{Parity:}
a parity function is a Boolean function whose value is 1 if and only if the input vector has an odd number of 1s. $PAR_{n}$ refers to the parity function with n bits.
\end{definition}
The parity function of two inputs is also knows as the XOR function. The output of parity function is call the Parity bit.

First, from the Lupanov's Theorem we can observe a fact that every function can be $\epsilon$-approximated by a DNF of size $(1 - \epsilon)2^{n-1}$ and width $n$

Second, the theorem from Boppana-Hastad (1997) states that every DNF that can $.01-approximates$ Parity function has size at least $2^{n/16}$ and width at least $n/16$

These bounds are definitely not tight enough and not within the universal bound we discussed in Section 1. So, we want to show that the universal bound can apply to the approximation of parity functions. This can be easily done by the following steps:

\begin{itemize}
    \item Flip each 0 to X (a symbol to represent an unknown value) with a probability $\epsilon$.
    \item Add all the sub-cubes of dimension $\log\log{n}$ that cover only 1 and X.
\end{itemize}
Thus we know that there is a DNF of size $O(2^{n}/\log{n})$ that can $\epsilon-approximate$ the parity function.

However, the universal bound is still not tight enough. Next, we want to show a tighter upper bound of approximating parity functions.

\begin{theorem}
Parity functions can be $\epsilon$-approximated by a DNF of size $2^{(1-2\epsilon)n}$ and width $(1 - 2\epsilon)n$
\end{theorem}
Theorem 2.1 states the upper bound of a DNF that can approximate the Parity function $PAR_{n}$. Based on Theorem 2.1, the size of the DNF is within the universal bound that stated in Theorem 1.1.

Let $\epsilon = 1/4$, we can construct a DNF Approximator for $PAR_{n}$ and argue that the size of the DNF is $2^{n/2}$ and width is $n/2$. In order to do this, select an input x, which x has n bits. Then partition x into two equal-length part y and z,
\begin{equation}
    PAR(x) = PAR(y)\oplus PAR(z)
\end{equation}
Consider $f(x) = PAR(y) \lor PAR(z)$, then we know that $Pr[f(x) = PAR(x)] = 3/4$:
\begin{itemize}
    \item $PAR(x) = 1 \rightarrow f(x) = 1$
    \item $PAR(x) = 0 \rightarrow f(x) = 0$ with probability $1/2$
\end{itemize}
Since PAR(y) and PAR(z) have trivial DNFs of size $2^{n/2 - 1}$ and width $n/2$. We get the DNF with size $2^{n/2}$ and width $n/2$ that $\epsilon-approximates$ $PAR_{n}$. Hence the construction is finished.

\section{Approximating Monotone Boolean Functions}

\begin{definition}
For two bitstrings $x,y \in \{0,1\}^n$, we say that $\mathbf{x \preceq y}$ if $x_i \leq y_i$ for all $i \in [n]$.
\end{definition}

Monotone boolean functions are a large family of boolean functions with the requirement that $f(x) \leq f(y)$ for all $x \preceq y$. Given this requirement, a natural question is can we achieve a tighter bound than $O(2^n/\log n)$ for $\epsilon$-approximating monotone boolean functions? The answer is yes:
\begin{theorem}
Every monotone function $f: \{0,1\}^n \rightarrow \{0,1\}$ can be \\$\epsilon$-approximated by a monotone function $g$ of DNF size $2^{n-\Omega_\epsilon(\sqrt{n})}$, satisfying $g(x) \leq f(x)$ for all $x \in \{0,1\}^n$.
\end{theorem}

This is proven using two lemmas. They will need a few definitions first.
\begin{definition}
A \textbf{k-regular} DNF is a DNF where all of its terms have a width of $k$. A \textbf{regular} DNF is a DNF that is k-regular for some k.
\end{definition}

\begin{definition}
A \textbf{lower $\epsilon$-approximator} for a function $f$ is an \\$\epsilon$-approximator $g$ such that $g(x) \leq f(x)$ for all $x$.
\end{definition}

\begin{definition}
Let $f$ be a boolean function and $k \in [n]$. The \textbf{density} of $f$ at level $k$ is defined as $\mu_k(f) := \Pr_{\parallel x\parallel = k}[f(x)=1]$. Note: if $f$ is monotone, then $\mu_k(f) \geq \mu_{k-1}(f)$.
\end{definition}

\begin{lemma} 
For any $\epsilon > 0$, every monontone function $f$ is $\epsilon$-close to the disjunction $g$ of monotone DNFs, $g(x) = g_1(x) \vee \dots \vee g_t(x)$, where 
\begin{enumerate}
    \item $t \leq 2/\epsilon$
    \item each $g_i$ is $k_i$-regular \\for some $k_i \in \Big[(n/2) - \sqrt{nln(4/\epsilon)/2}, (n/2) + \sqrt{nln(4/\epsilon)/2}\Big]$
    \item the DNF size of $g_i$ is at least $(\epsilon/2) {\binom{n}{k_i}}$
    \item $g(x) \leq f(x)$ for all $x \in \{0,1\}^n$
\end{enumerate}
\end{lemma}

\textbf{Proof of Lemma 3.2}

Set $l := \sqrt{\frac{n \ln (4/\epsilon)}{2}}$.

For $k \in [n]$, define $f_k(x) := \vee \{T_x : \parallel x\parallel = k \text{ and } T_x \text{ is a minterm of } f\}$. Where $T_x(y)=1$ iff $y \succeq x$.

By the Chernoff bound, $\Pr_x[|\parallel x\parallel - \frac{n}{2}| \geq l] \leq \frac{\epsilon}{2}$. 

So, $f^*(x) := f_{\frac{n}{2}-l}(x) \vee \dots \vee f_{\frac{n}{2}+l}(x)$ is a lower $\frac{\epsilon}{2}$-approximator of $f$.

By the triangle inequality, it suffices to show a $g$ that is $\frac{\epsilon}{2}$-close to $f^*$ which satisfies all four conditions.

We set the output of the following algorithm to $g$:

\begin{algorithm}[H]
    \For{$k \in \{\frac{n}{2}-l, \frac{n}{2}+l\}$}{
        \If{$\Pr_{\parallel x\parallel = k}[T_x \text{ is a minterm of } f^*] < \frac{\epsilon}{2}$}{
            \text{Set }$f^*(x) =0$\text{ for all }$T_x$\text{ where }. $\parallel x\parallel = k$
        }
    }
\end{algorithm}

Intuitively, this algorithm outputs a function $g$ that is equal to $f^*$ except with outputs set to $0$ if less than a $\frac{\epsilon}{2}$ fraction of the inputs at a given layer define a minterm $T_x$ in $f^*$.

The next steps will demonstrate that $g$ fulfills all of the conditions.

\textbf{$g$ is a $\epsilon$-approximator of $f$: } At any given layer only a $\frac{\epsilon}{2}$ fraction of the inputs are altered, meaning $g$ is $\frac{\epsilon}{2}$-close to $f^*$, making $g$ a lower $\epsilon$-approximator of $f$.

\textbf{Condition 4: } The only edit the algorithm makes is setting $f^*(x)=0$, so $g(x) \leq f^*(x)$ for all $x$, meaning $g(x) \leq f(x)$. 

\textbf{Condition 2: } Define each $g_i(x) := \vee \{T_x : \parallel x\parallel = k_i \text{ and } T_x \\ \text{ is a min term of } g\}$. Thus, one can also write $g$ as $g(x) = g_1(x) \dots g_t(x)$.

Each $g_i$ is $k_i$-regular since each $T_x$ has width $\parallel x \parallel$. Also, $g_i$ only has weight when $k_i \in [\frac{n}{2}-l, \frac{n}{2}+l]$.

\textbf{Condition 3: } Each $g_i$ only has weight when a $>\frac{\epsilon}{2}$ fraction of $T_x$ are in $g$ for $\parallel x \parallel = k_i$. So the size of each $g_i$ is $\geq \frac{\epsilon}{2} {\binom{n}{k_i}}$.

\textbf{Condition 1: } We will prove that $\mu_{k_i} (g_1 \vee \dots \vee ) \geq \frac{i\epsilon}{2}$ for all $i \in [t]$. This implies condition 1 because 
\begin{align*}
    \mu_{k_t}(g_1 \vee \dots \vee g_t) &\leq 1 \\
    \frac{t\epsilon}{2} &\leq 1 \\
    t &\leq \frac{2}{\epsilon}
\end{align*}

Assume without loss of generality that $k_1 < \dots < k_t$.

Suppose $\mu_{k_i}(g_1 \vee \dots \vee g_i) \geq \frac{i\epsilon}{2}$ for some $i < t$.

Because the $g$'s are monotone functions, $\mu_{k_{i+1}}(g_1 \vee \dots \vee g_i) \geq \mu_{k_i}(g_1 \vee \dots \vee g_i) \geq \frac{i\epsilon}{2}$.

To find $\mu_{k_{i+1}}(g_1 \vee \dots \vee g_{i+1})$, note that the terms of $g_{i+1}$ are disjoint from $g_1 \vee \dots \vee g_i$ because all of the $g_{i+1}$ terms have width of ${k_{i+1}}$. Thus:

\begin{align*}
    \mu_{k_{i+1}}(g_1 \vee \dots \vee g_{i+1}) &= \mu_{k_{i+1}}(g_1 \vee \dots \vee g_i) + \mu_{k_{i+1}}(g_{i+1}) \\
    &\geq \frac{i\epsilon}{2} + \frac{\epsilon}{2} = (i+1) \frac{\epsilon}{2}
\end{align*}

Therefore, $\mu_{{k_{i+1}}}(g) \geq \frac{t\epsilon}{2}$.

This concludes the proof of Lemma 3.2.

\begin{lemma}
Let $f$ be a regular monotone function. For every $\epsilon > 0$ there exists a montone DNF $g$ of size $2^{n-\Omega(\epsilon\sqrt{n} - log(n))}$ that is a lower $\epsilon$-approximator for $f$.
\end{lemma}

\textbf{Proof of Lemma 3.3}

We may assume that $\epsilon \geq \frac{C \log n}{\sqrt{n}}$ because otherwise, the size would be $2^{n-\Omega(1)}$ which is already true by Theorem 1.2.

Let $f$ be a k-regular monotone function for some $k \in [n]$.

Our approximator $g$ will be disjunctions of some terms $T_y$ where each $y \in f^{-1}(1)$ and $T_y(x) = 1$  for all $x \succeq y$. This construction makes $g$ a lower approximator for $f$.

This proof will first divide the inputs by hamming weight and then reduce the problem to only consider a smaller subset of the inputs. Note that the hamming weight of a uniformly distributed input is the same as sampling from a binomial distribution. 

By the Chernoff bound, $\Pr_x [\parallel x \parallel \geq \frac{n}{2} + t\frac{\sqrt{x}}{2}] \leq e^{-t^2/2}$. Setting $t=\sqrt{2 \ln (\frac{3}{\epsilon})}$, we get $\Pr_x\Big[\parallel x \parallel  \geq \frac{n}{2} + \sqrt{\frac{n\ln(3/\epsilon)}{2}}\Big] \leq \frac{\epsilon}{3}$.

By the anti-concentration of the Binomial, for an interval $I \subseteq [0,n]$ of width at most $\epsilon \sqrt{n}$, we have $\Pr_x [\parallel x\parallel  \in I] \leq 2\epsilon$. Using an interval of $[k, k+\frac{\epsilon\sqrt{n}}{6}]$, $\Pr_x\big[\parallel x\parallel  \in [k, k+\frac{\epsilon\sqrt{n}}{6}]\big] \leq \frac{\epsilon}{3}$.

Notice that if $\parallel x\parallel  < k$, then $f(x) = 0$ because $f$ is k-regular. Also, if our approximator outputs $0$ whenever $\parallel x\parallel  \geq \frac{n}{2} + \sqrt{\frac{n\ln(3/\epsilon)}{2}}$ or $\parallel x\parallel  \in [k, k+\frac{\epsilon\sqrt{n}}{6}]$, it will be wrong with an extra probability of at most $\frac{2\epsilon}{3}$. So for the remaining interval, $A:=\{x \in \{0,1\}^n : \parallel x\parallel \in[k+\frac{\epsilon\sqrt{n}}{6}, \frac{n}{2} + \sqrt{\frac{n\ln (3/\epsilon)}{2}}\}$, $\Pr_{x\sim A}[g(x) \neq f(x)] \leq \frac{\epsilon}{3}$. If this holds, $g$ will be a lower $\epsilon$-approximator of $f$ because 

\begin{align*}
\Pr_x[f(x) \neq g(x)] &\leq 
1*\Pr_x[\parallel x\parallel  \in [k, k+\frac{\epsilon\sqrt{n}}{6}]] + \\
& 1*\Pr_x\Big[\parallel x\parallel  \geq \frac{n}{2} + \sqrt{\frac{n\ln(3/\epsilon)}{2}}\Big] +\\ 
&\frac{\epsilon}{3}*\Pr_x[x \in A] + 0*\Pr_x[\parallel x \parallel < k] \\
&\leq 3*\frac{\epsilon}{3} = \epsilon
\end{align*}

For $l \in [n-k]$, we say $S_l$ is the set of 1-inputs with hamming weight $k+l$.

If for each $l \geq \frac{\epsilon\sqrt{n}}{6}$, there exists a monotone DNF $g_l$ satisfying:

\begin{enumerate}[(i)]
    \item minterms of $g_l$ are of the form $T_y$ for $y \in S_{l/2}$.
    \item size of $g_l = O(2^{n-l/2}) \leq 2^{n-\Omega(\epsilon\sqrt{n})}$
    \item $\Pr_{x\in S_l} [g_l(x) = 0]\leq \frac{\epsilon}{3}$
\end{enumerate}

Then by setting $g$ to be the disjunction of all $g_l$ where $k+l \in [k+\frac{\epsilon\sqrt{n}}{6}, \frac{n}{2} + \sqrt{\frac{n \ln (3/\epsilon}{2}}]$, the size of $g$ will be at most $n*2^{n-\Omega(\epsilon\sqrt{n})} \leq 2^{n-\Omega(\epsilon\sqrt{n} - \log n)}$. And by (iii), $\Pr_{x\in A}[g(x) \neq f(x)] \leq \frac{\epsilon}{3}$, which would complete the proof.

We generate each $g_l$ by sampling from the following distribution $\mathcal{D}$. For each $y \in S_{l/2}$, include $T_y$ as a minterm of $g_l$ with probability $p:=2^{-l/2}$. Now we will show that this construction obeys the three conditions.

(i) is satisfied by the definition.

(ii) The size of the term $g_l$ follows a binomial distribution, so $E_{g_l \sim \mathcal{D}} [g_l $ size$] = p* |S_{l/2}| < p2^n = 2^{n-l/2}$. By Markov's inequality, set $a=3*2^{n-l/2}$ and 
\begin{align*}
    \Pr[X \leq a] \geq 1- \frac{E[X]}{a} \\
    \Pr[g_l \text{ size} \leq 3*2^{n-l/2}] \geq \frac{2}{3}
\end{align*}

So with positive probability, the size is upper bounded by $O(2^{n-l/2})$. 

(iii) Take any $x\in S_l$. there must exist a $z \in S_0$ such that $z \prec x$ because the $z$ will correspond to a minterm in $f$ which all have hamming weight $k$. Also, there are ${\binom{1}{1/2}}$ many $y \in S_{l/2}$ where $z \prec y \prec x$ because $x$ is $1$ on exactly $l$ bits, and $y$ is $1$ on exactly $\frac{l}{2}$ of its bits.

The probability of a $g_l$ sampled from $\mathcal{D}$ outputting $g_l(x) = 0$ is at most the probability of picking none of the $y \in S_{l/2}$ described above. So, 

\begin{align*}
    \Pr_{g_l \sim \mathcal{D}}[g_l(x) = 0] &\leq (1-p)^{\Theta(2^l/\sqrt{l})} \\
    &= e^{-\Omega(2^{l/2}/\sqrt{l})} \\
    &< e^{-\Omega(2^{\epsilon\sqrt{n}}/12)/\sqrt{n}}  < \frac{\epsilon}{9}
\end{align*}

Therefore, $E_{g_l \sim \mathcal{D}} [\Pr_{x \in S_l} [g_l(x) = 0]] < \frac{\epsilon}{9}$. And by Markov's inequality, $\Pr_{g_l \sim \mathcal{D}} [\Pr_{x \in S_l}[g_l(x) = 0] \leq \frac{\epsilon}{3}] \geq 1 - \frac{\epsilon}{9}*\frac{3}{\epsilon} = \frac{2}{3}$.

So in all, the probability that $g_l$ is the correct size (ii) and is correct on enough inputs (iii) is the inverse of:
\begin{align*}
    \Pr[\text{(ii) not satisfied} \vee \text{(iii) not satisfied}]  
    \\ \leq\Pr[\text{(ii) not satisfied}] + \Pr[\text{(iii) not satisfied}]  \\
    = \frac{1}{3} + \frac{1}{3} = \frac{2}{3}
\end{align*}

So the probability that both (ii) and (iii) are satisfied is $\geq 1-\frac{2}{3} = \frac{1}{3}$.

Thus, there is a positive probability that $g_l$ satisfies all three conditions which means it exists. 

This concludes the proof of Lemma 3.3.

\subsection{Proof of Theorem 3.1}

By Lemma 3.2, every monotone $f$ has a lower $\epsilon$-approximator $g(x) = g_1(x) \vee \dots g_t(x)$ where $t \leq 2/\epsilon$ and each $g_i(x)$ is a regular monotone function. 

By Lemma 3.3, each $g_i$ has a lower $\frac{\epsilon}{2t}$-approximator $h_i$ of size \\ $2^{n-\Omega((\epsilon\sqrt{n}/t) - \log (n))}$.

By using the union bound, we get:

\begin{align*}
    \Pr_x[g(x) \neq h(x)] = \Pr_x[h_1(x) \neq g_1(x) \vee \dots \vee h_1(x) \neq g_1(x)]     \leq t \frac{\epsilon}{2t} = \frac{\epsilon}{2}
\end{align*}

By using the triangle inequality, we get:

\begin{align*}
    \Pr_x[h(x) \neq f(x)] \leq \Pr_x[g(x) \neq f(x)] + \Pr_x[h(x) \neq g(x)] = \frac{\epsilon}{2} + \frac{\epsilon}{2} = \epsilon
\end{align*}

And so $h$ is a lower $\epsilon$-approximator of $f$ with size $\leq t * 2^{n-\Omega((\epsilon\sqrt{n}/t) - \log(n)} = 2^{n-\Omega_\epsilon(\sqrt{n})}$. 

This concludes the proof of Theorem 3.1.

\section{Approximating Majority Function with DNF}
Majority Function $Maj_{n}$ is one of the most common Boolean functions that used in computer science. $Maj_{n}$ can also be approximated by DNF. We are interested in what is the upper bound of the size of the DNF that can approximate Majority Function, and whether this upper bound is inside the universal bound that we proved in Section 1.
\newline
First, what is the upper bound of the DNF that approximates Majority Function $Maj_{n}$?
\begin{theorem}
The DNF of size $2^{O(\sqrt{n}/\epsilon)}$ can $\epsilon-approximates$ Majority on n bits $Maj_{n}$
\end{theorem}
In order to prove this theorem, we give a construction of a DNF whose size is $2^{O(\sqrt{n}/\epsilon)}$ and show this DNF approximates $Maj_{n}$. The construction is inspired by the random DNF construction of Talagrand.

\begin{theorem}
For all $\epsilon \ge 1/\sqrt{n}$, there is a DNF of width $w = 1/\epsilon \sqrt{n}$ and size $(ln2)2^{w}$ that $O(\epsilon)-approximates$ Majority with n bits $Maj_{n}$
\end{theorem}

To prove this theorem, let D be a randomly chosen DNF with $(ln2)2^{w}$ terms, where each term is chosen by
picking w variables independently with replacement. Then, to prove Theorem 4.2, we can just show that
\begin{equation}
    E_{D}[Pr_{x}[D(x) \neq Maj(x)]] \le O(\epsilon)
\end{equation}
This is equivalent with showing
\begin{equation}
    E_{x}[Pr_{D}[D(x) \neq Maj(x)]] \le O(\epsilon)
\end{equation}
Let $t\in [-\sqrt{n}, \sqrt{n}]$, given a string $x \in {0, 1}^{n}$, the fraction of 1's in x is $1/2 + 1/2(t/\sqrt{n})$ because of the Central Limit Theorem. Since $Maj(x) = 1$ if and only if $t > 0$, by construction, $Pr_{D}[D(x)=1]$ only depends on t. So we have:
\begin{equation}
    Pr_{D}[D(x)=1] = 1 - (1 - 2^{-2}(1+t/\sqrt{n})^{w})^{(\ln 2)2^{w}}
\end{equation}
Now, in order to prove Theorem 4.2, it is sufficient to show that
\begin{center}
    $E_{x}[(1 - 2^{-w}(1 + t/\sqrt{n}))^{(\ln 2)2^{w}} | t > 0] \le O(\epsilon)$
\end{center}
\begin{center}
    and
\end{center}
\begin{center}
    $E_{x}[1 - (1 - 2^{-w}(1 + t/\sqrt{n}))^{(\ln 2)2^{w}} | t > 0] \le O(\epsilon)$
\end{center}
Note the fact that $(1 - x)^{y} \le \exp{(-x y)}$ and $(1 - x)^{y} \ge 1 - x y$, using $w = 1/\epsilon\sqrt{n}$, we can get
\begin{center}
    $(1 - 2^{-w}(1 + t/\sqrt{n}))^{(\ln 2)2^{w}} \le (1/2)^{(1 + t/\sqrt{n})^{w}} \le (1/2)^{1 + t/\epsilon}$
\end{center}
\begin{center}
    and
\end{center}
\begin{center}
    $1 - (1 - 2^{-w}(1 + t/\sqrt{n}))^{(\ln 2)2^{w}}
    \le (\ln 2)\exp (wt / \sqrt{n}) = (\ln 2)\exp (t / \epsilon)$
\end{center}

Then there is only one thing remains to show, which is $E_{x}[\exp (-|t|/\epsilon)] \le O(\epsilon)$. Note the fact that for each $i = 0, 1, 2, ...$,
\begin{center}
    $Pr[|t| \in [2^{i}\epsilon, 2^{i+1}\epsilon]] = Pr[|Normal(0, 1)| \in [2^{i}\epsilon, 2^{i+1}\epsilon]] + O(1/\sqrt{n})$
\end{center}
 where $O(1/\sqrt{n}$ is negligible since $\epsilon \ge 1/\sqrt{n}$. Using $Pr[|t| \in [0, \epsilon]] \le O(\epsilon)$, we get
\begin{equation}
    E_{x}[\exp (-|t|/\epsilon)] \le O(\epsilon) + \sum_{n=1}^{\infty} \exp (-2^{n})O(2^{i}\epsilon) \le O(\epsilon)
\end{equation}
Hence, we finished the proof.
\newline
After we proved the upper bound of DNF that approximates $Maj_{n}$, we are going to show that this upper bound is inside the universal upper bound $O_{\epsilon}(2\textsuperscript{n}/log\emph{n})$.
\newline
As we proved in this section, the DNF that approximates Majority has an upper bound of size $2^{O(\sqrt{n}/\epsilon)}$.
Let $a = 2^{\sqrt{n}}$, $b = 2^{n}/\log(n)$, show that $a/b \le 1$.
\begin{equation}
    a/b = 2^{\sqrt{n}}/(2^{n}/\log(n)) = 2^{\sqrt{n}} log(n)/2^{n} = log(n)/2^{\sqrt{n}}
\end{equation}
Clearly, when n becomes larger, $log(n)/2^{\sqrt{n}}$ becomes smaller and $log(n)/2^{\sqrt{n}} \le 1$.
This indicates that $2^{O(\sqrt{n}/\epsilon)} \le 2\textsuperscript{n}/log\emph{n}$. Hence the upper bound of DNF that approximates Majority is inside the universal bound of DNF approximation.

\section{Conclusion}
The Disjunctive Normal Form is a strong Boolean function that can be used to approximate other Boolean functions. The size of DNF that for approximation is always within a specific universal upper bound. By showing the DNF approximation of Parity function, Monotone function and Majority function in Section 2, 3 and 4. We can observe that the universal bound does apply to these well-known Boolean functions. In addition, the universal bound typically is not strict enough for some specific Boolean functions. For example, the Parity function and Majority both have a tighter upper bound. However, those upper bounds are specific to the Boolean function, which cannot be applied to Boolean functions in general.

\newpage
\section{References}
\begin{itemize}
    \item Eric Blais, Li-Yang Tan, \textit{Approximating Boolean Functions with Depth-2 Circuits.}, SIAM J. Comput. 44(6): 1583-1600 (2015) Preliminary version: Electronic Colloquium on Computational Complexity (ECCC) 20: 51 (2013) 
    
    \item Eric Blais, Johan Hastad, Rocco A. Servedio, and Li-Yang Tan, \textit{On DNF approximators for monotone Boolean functions.} (2018)
    
    \item O’Donnell, R., Wimmer, K. (n.d.)., \textit{Approximation by DNF: Examples and Counterexamples. }, Automata, Languages and Programming Lecture Notes in Computer Science, 195–206. doi: 10.1007/978-3-540-73420-8-19 (2006)
    
    \item Oleg Lupanov., \textit{Implementing the algebra of logic functions in terms of constant depth formulas in the basis  $\And, \lor, \neg$}, Dokl. Ak. Nauk. SSSR, 136:1041–1042 (1961)
    
    \item Aleksej Dmitrievich Korshunov. \textit{O slozhnosti kratchaıshikh dizyunktivnykh normalnykh form sluchanykh bulevykh funktsiı.} Metody Diskretnogo Anal, 40:25–53 (1983)
    
    \item S. E. Kuznetsov. \textit{O nizhneı otsenke dliny kratchaısheı dnf pochti vsekh bulevykh funktsiı.} Veroyatnoste Metody Kibernetiki, 19:44–47 (1983)
    
    \item M. Talagrand. \textit{How much are increasing sets positively correlated?} Combinatorica, 16(2):243–258, (1996)
    
    \item Eric Blais, Li-Yang Tan. \textit{Approximating Boolean functions with depth-2 circuits [PowerPoint slides]}.Retrieved from\newline simons.berkeley.edu/sites/default/files/docs/585/tanslides.pdf (2013)
    
    \item Eric Blais, Li-Yang Tan. \textit{Approximating functions with DNFs [PowerPoint slides]}. Retrieved from http://grigory.us/files/theory/eric.pdf
    
\end{itemize}

\end{document}